\documentclass[
 reprint,
 amsmath,amssymb,
 aps,
 prl
]{revtex4-2}

\usepackage{graphicx}% Include figure files
\graphicspath{{./Images/}}
\usepackage{dcolumn}% Align table columns on decimal point
\usepackage{bm}% bold math
\usepackage{braket}

\usepackage[english]{babel}
\usepackage[T1]{fontenc}
\usepackage[utf8]{inputenc}

\begin{document}

\title{Riemann Rarefaction Waves in a Strongly Interacting Fermi Gas}

\author{Eric A. Wolf$^1$}

\author{Martin Zwierlein$^1$}

\affiliation{
 $^1$MIT-Harvard Center for Ultracold Atoms, Research Laboratory of Electronics, and Department of Physics,
 Massachusetts Institute of Technology, Cambridge, Massachusetts 02139, USA}

\date{\today}

\begin{abstract}
    We investigate the expansion of a homogeneous, strongly interacting Fermi gas released into vacuum in a ``shock tube'' geometry. At unitarity, where the gas is scale invariant and nearly inviscid, we find that the resulting rarefaction wave dynamics are self-similar and in excellent agreement with Riemann's solution of the Euler equation for all temperatures probed. Probing interactions away from unitarity within the BEC-BCS crossover, we observe increasing deviations from the Riemann solution as viscosity increases. However, even on the BCS side, where the sound diffusivity is increased twenty-fold, self-similarity is still approximately preserved. This may reflect how 1D Navier-Stokes rarefaction flows approach Euler self-similar solutions at long times. Our work demonstrates the utility of strongly interacting Fermi gases for the study of nonlinear hydrodynamics in a highly controllable setting.
\end{abstract}

\maketitle

%%%%%%INTRO%%%%%%%%%

Understanding transport in strongly interacting Fermi systems is among the most challenging problems in many-body physics, encountered in contexts ranging from high-temperature superconductors~\cite{Keimer2015} to neutron stars~\cite{Schmitt2018} and the quark-gluon plasma~\cite{Schaefer2009,Shuryak2017}. Collision rates on the scale of the Fermi energy render the notion of weakly interacting quasi-particles inapplicable, and with it the standard transport behavior of Fermi liquids. Still, in the long-wavelength limit, conservation laws ensure that the dynamics of these systems are governed by hydrodynamics, and only the values of transport coefficients, e.g. viscosity and thermal conductivity, are challenging to predict.

Strongly interacting Fermi gases of ultracold atoms provide a highly controllable platform to test theories of transport~\cite{Giorgini2008,Ketterle2008,Zwerger2012,Zwierlein2016a,Pitaevskii2016,Krinner2017}. Even above the critical temperature for superfluidity, these gases realize a ``perfect fluid''~\cite{Schaefer2009} with low, quantum limited diffusion of spin~\cite{Sommer2011,Luciuk2017}, momentum~\cite{Cao2011}, sound~\cite{Patel2020,Huang2025} and heat~\cite{Xi2022,Yan2024,Xiang2024}, all on the scale of $\hbar/m$, where $m$ is the atomic mass.

A further simplification arises at the point of unitarity limited interactions, where the s-wave scattering length $a$ diverges. Here, the system is scale invariant; this implies a vanishing bulk viscosity  and a universal form $P = P_0\, f(s)$ for the equation of state, with $P_0 = 2/5\, n E_F$ the pressure of a non-interacting Fermi gas at zero temperature, $E_F$ the Fermi energy, and $f(s)$ a universal function of the entropy per particle $s{=}S/N k_B$~\cite{Ku2012}. As such, strong interactions considerably simplify the hydrodynamic description, which is especially useful for nonlinear hydrodynamics.

The low dissipation of the unitary gas enabled the use of the Euler equation~\cite{Menotti2002}, i.e. inviscid hydrodynamics, to understand the anisotropic expansion -- elliptic flow~\cite{Schaefer2009} -- in the original demonstration of strongly interacting Fermi gases~\cite{Ohara2002}. Similar approaches were used in work on collective modes~\cite{Schaefer2012, Tey2013, Hou2013, Kuznetsov2020, Hou2021}, shock waves~\cite{Joseph2011}, and expansion experiments probing the effects of viscosity~\cite{Cao2011, Joseph2015} and scale invariance~\cite{Elliott2014Sym, Wang2024}.

These previous probes of nonlinear hydrodynamics were carried out in inhomogeneous gases confined in a non-uniform external trapping potential, which implies an inhomogeneous entropy distribution and complications related to transport coefficients varying in space and becoming ill-defined in the low-density regions.

%%%%%BEGIN EXPERIMENT%%%%%

\begin{figure}
    \includegraphics[width = \linewidth]{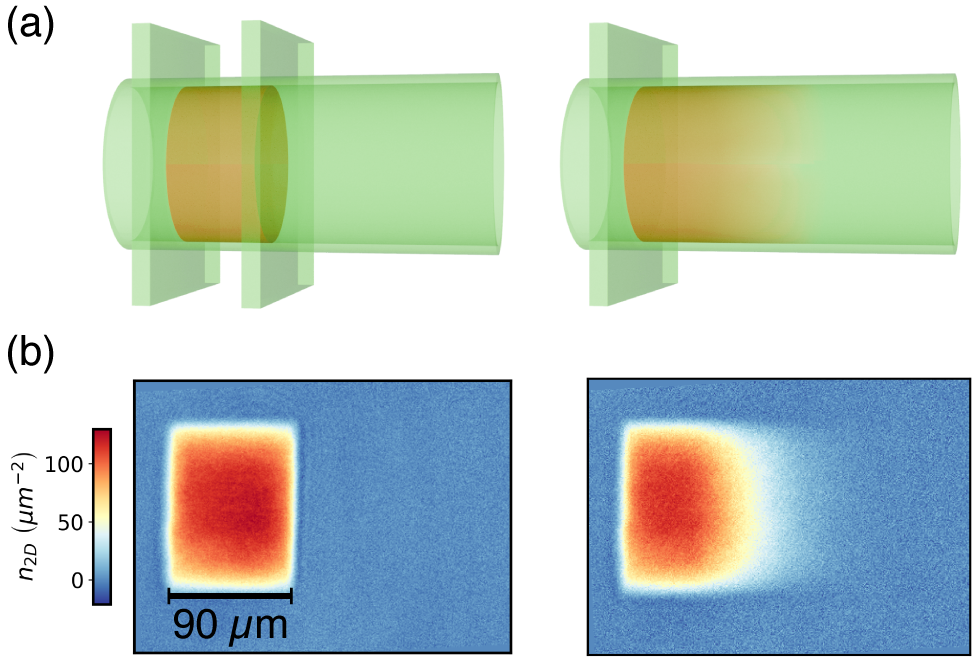}
    \caption{Experimental protocol. (a) A schematic 3D render of our ``shock tube'' geometry before (left) and after (right) gas release. (b) Typical two-dimensional density profiles for expansion time $t = 0$ ms (left) and $t = 2.5$ ms (right).}
    \label{Figure_1}
\end{figure}

In this work, we perform a classic nonlinear hydrodynamics experiment - the expansion of a gas into vacuum - using a homogeneous Fermi gas initially confined in a cylindrical box acting as a ``shock tube'' (see Fig.~\ref{Figure_1})~\cite{Mukherjee2017}. In this one-dimensional geometry, the inviscid hydrodynamic equations precisely realize a Riemann problem~\cite{Riemann1860}, an initial value problem for the Euler equations with a step-function initial condition. The resulting dynamics are self-similar, and we indeed find that our experimental data for all expansion times and even varying initial temperatures collapses onto a single curve: the rarefaction wave solution to Riemann's problem.

The Fermi gas consists of an ultracold, degenerate balanced spin mixture of the first and third lowest hyperfine states of $^6$Li, $\ket{1}$ and $\ket{3}$, prepared at the $\ket{1}-\ket{3}$ Feshbach resonance near 690 G~\cite{Bartenstein2005, Schunck2008, Zurn2013}. The confining cylindrical box (length $L{=}90\,\rm \mu m$, diameter $130\,\rm \mu m$) is created with blue-detuned laser beams, a ring-shaped beam for the radial confinement and two green light sheets as end caps. A typical density per spin species is $n_\mathrm{3D} {\sim} 1\ \mu \mathrm{m}^{-3}$, corresponding to a Fermi energy $E_F {\sim} h \cdot 10$ kHz. The initial speed of sound $c_0$ in the gas is measured from the frequency of the lowest symmetric sound resonance in the box~\cite{Patel2020}, with a typical value of $c_0 {\sim} 20$ mm/s at unitarity.

At time $t{=}0$, we suddenly turn off one of the end caps, and the gas expands into vacuum along the symmetry ($x$) axis of the cylinder. After a variable expansion time $t$, a polarization rotation image yields the radially-integrated density profile $n(x,t)$, or equivalently mass density $\rho = m n$, which we divide by the pre-expansion density to obtain the normalized mass density $\tilde{\rho}$.

\begin{figure}
    \includegraphics[width = \linewidth]{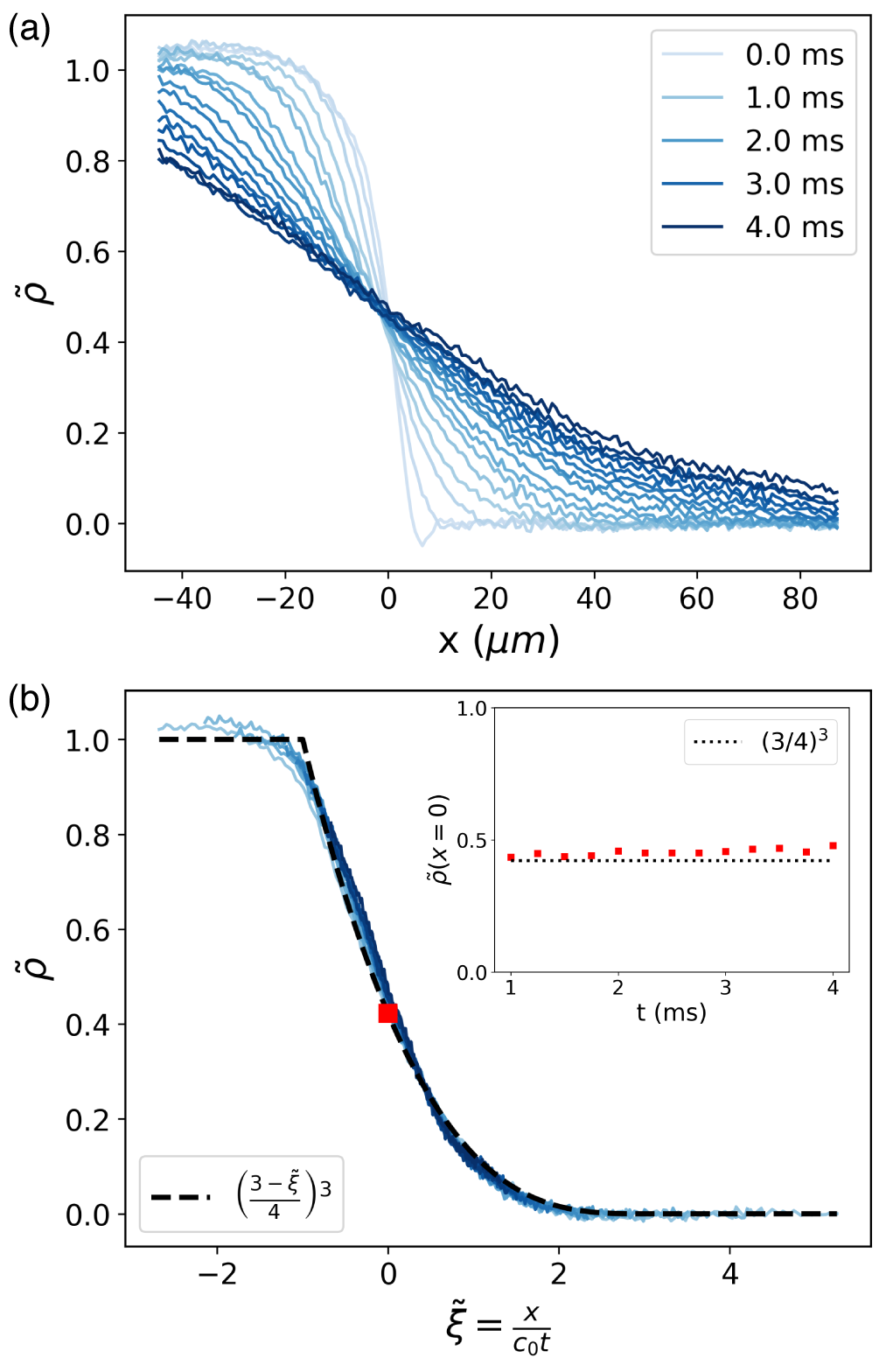}
    \caption{Self-similar expansion. (a) A collection of normalized densities vs. absolute position in the vicinity of $x = 0$, the original position of the box wall. There are a total of 17 traces, corresponding to $0$ ms to $4$ ms expansion in steps of $0.25$ ms; for clarity, only some traces are labeled. (b) A subset of the data from (a), with expansion times from $t = 1.0$ ms to $t = 4.0$ ms, plotted as a function of the reduced coordinate $\tilde{\xi} \equiv \frac{x}{c_0 t}$. The black dashed curve is Eq. \ref{Rho_Equation_Final}. Inset: The normalized density $\tilde{\rho}$ at $x {=} 0$ for different times; a dotted line marks the predicted value $\left(\frac{3}{4}\right)^3$ from Eq.~\ref{Rho_Equation_Final}.}
    \label{Figure_2}
\end{figure}

Fig.~\ref{Figure_2}(a) shows $\tilde{\rho}$ profiles thus obtained from free expansion of the unitary Fermi gas at a temperature of $\frac{T}{T_F} = 0.15$. The initially sharp transition from finite to zero density at the position of the box wall at $x = 0$ gradually smoothens, with the characteristic length scale of the density profile increasing with time. It is notable that the various profiles cross at $x = 0$, a first hint of the self-similar nature of the dynamics. Strikingly, when space $x$ is rescaled by the expansion time $t$ in Fig.~\ref{Figure_2}(b), the curves for different expansion times all collapse onto a single profile.

\begin{figure*}
    \includegraphics[width = \linewidth]{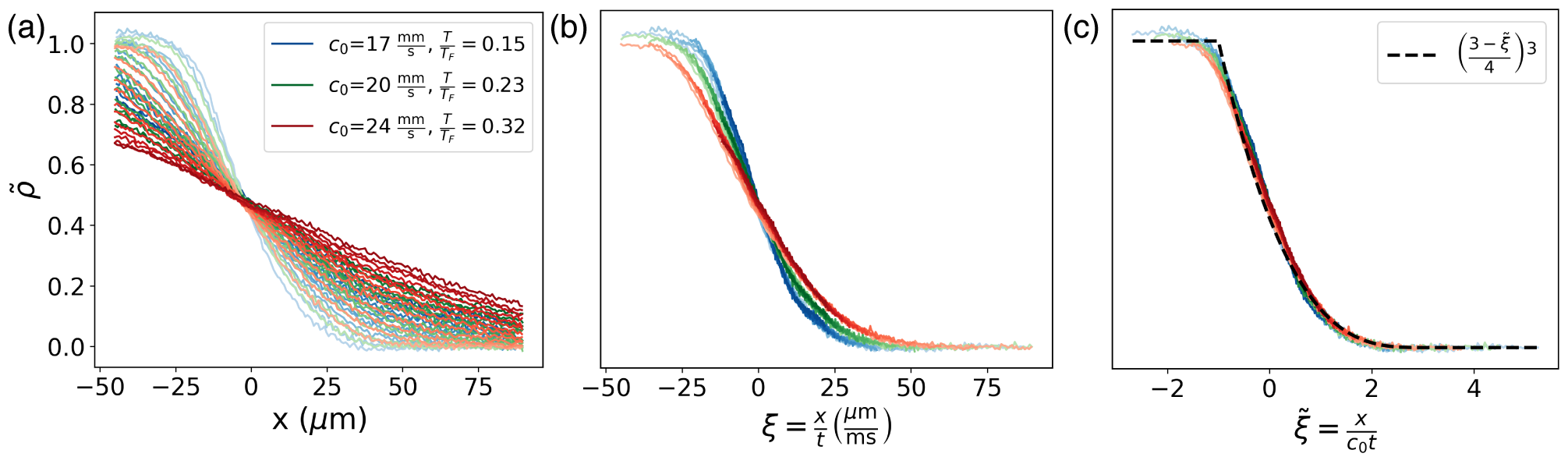}
    \caption{Expansion at different $c_0$. (a) A collection of normalized densities $\tilde{\rho}$ vs. x, for expansion times from 1.0 to 4.0 ms. Plotted in blue is the same data as Figure \ref{Figure_2}(a); green and red curves are analogous data obtained with different sound speeds. As in Figure \ref{Figure_2}, darker curves are at later times. (b) The same data as (a), plotted vs. $\xi = \frac{x}{t}$. (c) Again the same data, plotted vs. $\tilde{\xi} = \frac{x}{c_0 t}$, with each dataset normalized by its own $c_0$, as labeled. The dashed line is Eq. \ref{Rho_Equation_Final}.} 
    \label{Figure_3}
\end{figure*}

We may understand this collapse readily from the fact that the initial pressure $P$ and density $\rho$, which govern the 1D expansion, are quantities from which one cannot form a characteristic length scale - only a characteristic velocity scale ($\sqrt{P/\rho}$)~\cite{LandL}. This immediately implies that the dynamics must be self-similar, depending only on the ratio $\xi = \frac{x}{t}$. Viscosity $\eta$ would introduce a length scale $\eta /\sqrt{P \rho}$, but for the degenerate, unitary Fermi gas this is on the scale of the interparticle spacing, far below the length scale of hydrodynamic expansion $\sqrt{P/\rho}\cdot t$.

Following Riemann~\cite{Riemann1860}, we can now derive the profile of the expansion: a rarefaction wave. An inviscid fluid in a cylindrical ``pipe'', homogeneous in the radial direction, obeys the 1D continuity and Euler equations resulting from mass and momentum conservation~\cite{LandL}:
\begin{equation}
    \frac{\partial \rho}{\partial t} + \rho \frac{\partial v}{\partial x} + v \frac{\partial \rho}{\partial x} = 0
    \label{Euler_rho}
\end{equation}
\begin{equation}
    \frac{\partial v}{\partial t} + v \frac{\partial v}{\partial x} = - \frac{1}{\rho} \frac{\partial P}{\partial x}
    \label{Euler_v}
\end{equation}
where $v(x)$ is the 1D velocity field and $P(\rho,s)$ is the pressure, which depends on the local density $\rho$ and entropy per particle $s$. To close the system, energy conservation yields the continuity equation for $s$:
\begin{equation}
    \frac{\partial s}{\partial t} + v \frac{\partial s}{\partial x} = 0
    \label{Entropy_orig}
\end{equation}
Assuming that $s$ depends only on $\xi = x/t$, the entropy equation for $s(\xi)$ becomes $(v-\xi)s' = 0$, implying constant entropy. Under insentropic conditions, pressure changes are tied to density changes as $P' = \left.\frac{\partial P}{\partial \rho}\right|_s\rho' = c^2 \rho'$ with the speed of sound $c$. In particular, for the unitary gas, at any particular value of the constant entropy, the equation of state is polytropic $P \propto \rho^\gamma$ with $\gamma = 5/3$~\cite{Patel2020}, and so $c^2 = \gamma P/\rho$ at all temperatures.

The continuity and Euler equations then yield $(v - \xi)^2 = c^2$ or $\xi = v - c$. The sign choice corresponds to rarefaction flow towards $\xi>0$, and the continuity equation simplifies to $c \rho' = - \rho v'$. As $c$ itself is a function of $\rho$, for any given value of the constant entropy, one can solve

\begin{equation}
    v = -\int \frac{c}{\rho} \mathrm{d} \rho \implies v = \frac{2}{\gamma - 1} (c_0 - c)
    \label{V_C_Equation_Free}
\end{equation}
where the former result is general, and the latter holds for any system with a polytropic equation of state, $P \propto \rho^\gamma$. The integration constant $c_0 = c(\rho_0)$, the speed of sound at the density of the box before expansion, ensures the initial condition $v(x,t=0)=0$. Combining with $\xi = v-c$, and using the polytropic index $\gamma = 5/3$ for the isentropic unitary gas, we obtain the density profile
\begin{equation}
    \tilde{\rho} = 
        \left(\frac{3 - \tilde{\xi}}{4}\right)^3, \  -1 \leq  \tilde{\xi} \leq 3
    \label{Rho_Equation_Final}
\end{equation}
where $\tilde{\xi} = \frac{x}{c_0 t}$, and, for $\tilde{\xi} {<} -1$ or $\tilde{\xi} {>} 3$, $\tilde{\rho} {=} 1$ or $0$, respectively.

Equation~\ref{Rho_Equation_Final}, shown as the black dashed line in Fig.~\ref{Figure_2}(b), is in excellent agreement with the experimentally observed self-similar density profile. We stress that this agreement is fit-free, using the independently measured speed of sound $c_0$ to from the expansion length $c_0 t$. In particular, Eq.~\ref{Rho_Equation_Final} predicts that $\rho(x = 0) = \left(\frac{3}{4}\right)^3$ at all times, in good agreement with the data (Fig. \ref{Figure_2}(b), inset). The small deviations between the theory prediction and the experimental data, in particular the rounding off of the sharp cusp at $\tilde{\xi} = -1$, can be explained by a combination of finite viscosity and nonzero wall thickness at $t = 0$~\cite{SI}.

We stress that Eq.~\ref{Rho_Equation_Final} is valid at all temperatures, as long as the assumption of inviscid hydrodynamics is fulfilled. To test self-similarity and the agreement with the Riemann solution, we heat the gas at unitarity. This increases the speed of sound~\cite{Patel2020} while preserving the polytropic dependence $P \propto \rho^\frac{5}{3}$ at the new, fixed entropy.

In Fig. \ref{Figure_3}, we present in blue, green, and red the normalized density curves for measurements of heated unitary gases with speeds of sound $c_0 = 16.6$, $19.9$, and $23.7$ $\mu$m/ms, and normalized temperatures of $\frac{T}{T_F} = 0.15$, $0.23$, and $0.32$, respectively.

Fig.~\ref{Figure_3}(a) displays the profiles against position $x$; as expected, the curves cross at $x = 0$ for all $t$ and $c_0$, but are otherwise widely spread. When this data is plotted as a function of the time-normalized variable $\xi = \frac{x}{t}$ in Fig.~\ref{Figure_3}(b), each colored dataset collapses onto a self-similar curve which is visibly distinct for each $c_0$ value. Then, when we further normalize to $\tilde{\xi} = \frac{x}{c_0 t}$ using the measured values of $c_0$, we see that all datasets collapse onto Eq.~\ref{Rho_Equation_Final} (Fig. \ref{Figure_3}(c)) - a striking demonstration that this profile describes the rarefaction dynamics of the unitary gas over a significant range of temperatures. 

%%%% VISCOSITY ENTERS THE GAME %%%%

So far we have focused on the unitary Fermi gas. We may now freely vary interactions across the Feshbach resonance to test the limits of self-similarity and inviscid Euler dynamics~\cite{Liao2026}. Interactions affect the problem in two major ways. First, the equation of state changes dramatically with interaction parameter $\left(k_F a\right)^{-1}$, with the system undergoing the crossover from a weakly interacting Bose gas of tightly bound molecules for $a>0$ to a weakly interacting attractive Fermi gas for $a<0$.
In either limit, we expect to recover a polytropic equation of state: In the limit of molecular Bose-Einstein condensation, $P \propto \rho^2$ so $\gamma = 2$, while for a weakly interacting Fermi gas we recover $\gamma = 5/3$. For intermediate interaction strengths, the equation of state is not polytropic, although an effective polytropic index $\gamma$ has been usefully defined in the crossover~\cite{Hu2004,Ketterle2008}.

The second major impact of interactions is on dissipation in the gas, affecting all transport properties from viscosity~\cite{Liao2026} and thermal conductivity to sound~\cite{Huang2025,Fernandez2025} and spin diffusion~\cite{Sommer2011}. As the scattering cross section $\propto a^2$ decreases away from unitarity, the mean free path $l = (n \sigma)^{-1}$ increases, and with it typical diffusivities such as the kinematic viscosity (momentum diffusion) $\nu = \eta/\rho \propto \bar{v} l$, where $\bar{v}$ is the rms speed.
Instead of inviscid Euler equations, hydrodynamics are now governed by the full Navier--Stokes equations. The presence of viscosity introduces a new length scale $\nu/c_0$ that should in principle break self-similarity.

As a measure of dissipation, we directly obtain the sound diffusivity $D$ in the BEC-BCS crossover from the width of the sound resonance used to determine the speed $c_0$~\cite{Patel2020}. From the combination of continuity, Navier--Stokes, and entropy flow equations, one can show that it is the sound diffusivity $D$ that sets the effective dissipation of density waves. $D$ contains contributions from viscosity $\eta$ as well as thermal conductivity, as thermal gradients couple to density gradients in compressible liquids or gases.

\begin{figure} 
    \includegraphics[width = \linewidth]{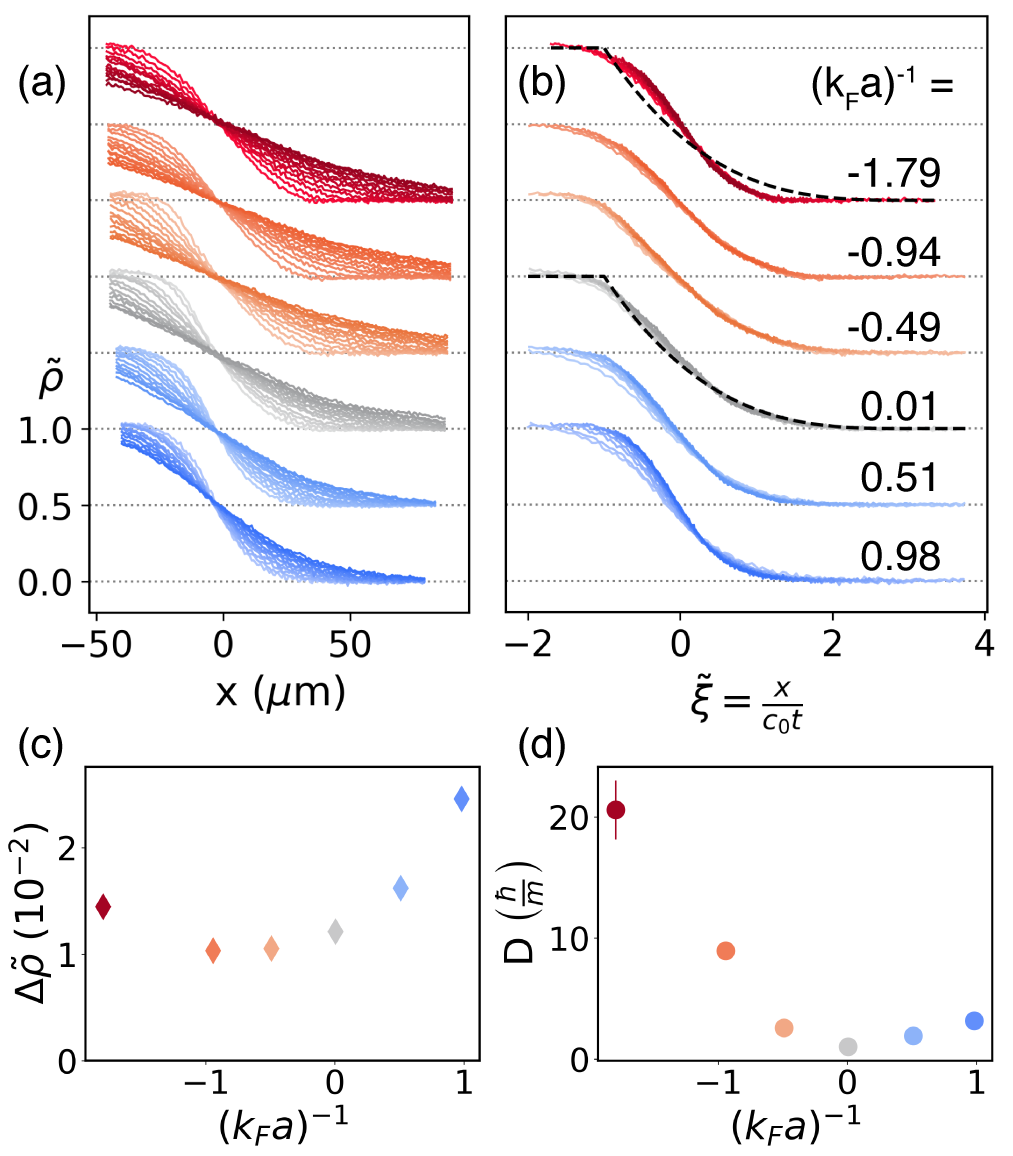}
    \caption{Expansion at different $k_F a$ values. (a) \& (b): Ridgeline plots of normalized density $\tilde{\rho}$ vs. (a) position $x$ and (b) $\tilde{\xi}$ for different initial values of the interaction parameter $\left(k_F a\right)^{-1}$. Displayed data is for expansion times from $t = 1.0$ ms (light) to $t = 4.0$ ms (dark). Adjacent curves are vertically offset by $0.5$. The gray curve, with $(k_F a)^{-1} = 0.01$, corresponds to the data plotted in Figure \ref{Figure_2}(b); the other curves are analogous, including the time range plotted, but are measured at different $(k_F a)^{-1}$ values. Dashed lines in (b) are Eq. \ref{Rho_Equation_Final}. (c) RMS deviations from self-similarity of the data plotted in (b). Deviations are calculated using a piecewise linear best-fit curve chosen for each dataset~\cite{SI}. (d) A plot of the sound diffusivities measured via box shaking resonances for each of the $(k_F a)^{-1}$ values plotted in (a) and (b).}
    \label{Figure_4}
\end{figure}

In Fig.~\ref{Figure_4}(a), we present the profiles $\tilde{\rho}$ obtained for box expansion performed at different interaction strengths in the BEC-BCS crossover. For consistency between the different datasets, the gas is prepared at unitarity, and subsequently the interactions are tuned adiabatically to the target value immediately before expansion.

Even away from unitarity, the curve collapse demonstrated in Fig.~\ref{Figure_2} largely persists, as shown in Fig.~\ref{Figure_4}(b). 
The RMS deviation $\Delta \tilde{\rho}$ of the density profiles from self-similarity (Fig.~\ref{Figure_4}c)~\cite{SI} is nearly constant as interactions are tuned into the BCS regime $(k_F a)^{-1} < 0$. This excellent collapse occurs despite the sound diffusivity $D$ increasing by a factor of 20. On the BEC side, self-similarity is less perfect at early times despite low dissipation; this may be due to the increased sensitivity of the more weakly interacting BEC to the finite sharpness of the box walls ~\cite{SI}.

Na\"ively, the persistence of self-similarity appears surprising. The argument for self-similarity relied on the absence of intrinsic length scales in Euler's equation. The kinematic viscosity $\nu$ or the sound diffusivity $D$ provide a new length scale, e.g. $D/c_0$, for which we obtain $\sim 1\ \mu$m at unitarity and $\sim 10\ \mu$m for the BCS dataset at $1/k_F a = -2$.

As $t \rightarrow \infty$, however, we expect the self-similar length scale $c_0 t$ to eventually satisfy $c_0 t \gg D/c_0$, such that viscous effects become negligible and the rescaled data $\tilde{\rho}(\tilde{\xi})$ again exhibits self-similarity. We can interpret this as a Reynold's number ${\rm Re} = c_0^2 t / D$ that grows in time as the system size becomes much larger than viscous length scales.
We have performed numerical simulations which suggest that, for $t \geq 1.0$ ms, the viscous contribution to non-self-similarity is sufficiently small as to be obscured by e.g. experimental imperfections~\cite{SI}.

Even with self-similarity preserved, the collapsed profiles $\tilde{\rho}(\tilde{\xi})$ clearly deviate from the Riemann solution Eq.~\ref{Rho_Equation_Final} in the deep BCS limit. As we expect to recover $P \propto \rho^{5/3}$ for the weakly interacting Fermi gas, we attribute this deviation to viscosity and thermal conductivity.

In conclusion, we observe self-similar rarefaction waves in the unitary Fermi gas during a 1D expansion into vacuum.
While self-similarity is remarkably robust as interactions are varied away from unitarity, increasing deviations from Riemann's solution point to the relevance of the full Navier--Stokes equations, including dissipative effects. It will be interesting to directly observe entropy production during expansion using thermography~\cite{Yan2024}, which would be a direct marker of non-Euler flow.
Our experimental platform serves as an ideal test bed for the study of highly nonlinear flows, and can be extended to launch e.g. controlled 1D shock waves~\cite{Joseph2011}.

We thank Keaton Burns, J\"orn Dunkel, Richard Fletcher, Victor Galitski, Leonid Levitov and Archisman Panigrahi for inspiring discussions, and Jens Hertkorn for a critical reading of the manuscript. This work was supported by the NSF through the Center for Ultracold Atoms and PHY-2513210, DOE (DESC0024622), and the DARPA APAQuS program. 

\bibliography{SelfSimilarRefs}

@PREAMBLE{
 "\providecommand{\noopsort}[1]{}" 
 # "\providecommand{\singleletter}[1]{#1}%" 
}

@article{Bartenstein2005,
  title = {Precise Determination of $^{6}\mathrm{Li}$ Cold Collision Parameters by Radio-Frequency Spectroscopy on Weakly Bound Molecules},
  author = {Bartenstein, M. and Altmeyer, A. and Riedl, S. and Geursen, R. and Jochim, S. and Chin, C. and Denschlag, J. Hecker and Grimm, R. and Simoni, A. and Tiesinga, E. and Williams, C. J. and Julienne, P. S.},
  journal = {Phys. Rev. Lett.},
  volume = {94},
  issue = {10},
  pages = {103201},
  numpages = {4},
  year = {2005},
  month = {Mar},
  publisher = {American Physical Society},
  doi = {10.1103/PhysRevLett.94.103201},
  url = {https://link.aps.org/doi/10.1103/PhysRevLett.94.103201}
}

@article{Cao2011,
author = {C. Cao  and E. Elliott  and J. Joseph  and H. Wu  and J. Petricka  and T. Schäfer  and J. E. Thomas },
title = {Universal Quantum Viscosity in a Unitary {Fermi} Gas},
journal = {Science},
volume = {331},
number = {6013},
pages = {58-61},
year = {2011},
doi = {10.1126/science.1195219},
URL = {https://www.science.org/doi/abs/10.1126/science.1195219}
}

@article{Elliott2014Sym,
  title = {Observation of Conformal Symmetry Breaking and Scale Invariance in Expanding {Fermi} Gases},
  author = {Elliott, E. and Joseph, J. A. and Thomas, J. E.},
  journal = {Phys. Rev. Lett.},
  volume = {112},
  issue = {4},
  pages = {040405},
  numpages = {5},
  year = {2014},
  month = {Jan},
  publisher = {American Physical Society},
  doi = {10.1103/PhysRevLett.112.040405},
  url = {https://link.aps.org/doi/10.1103/PhysRevLett.112.040405}
}

@article{Hou2013,
  title = {Scaling solutions of the two-fluid hydrodynamic equations in a harmonically trapped gas at unitarity},
  author = {Hou, Yan-Hua and Pitaevskii, Lev P. and Stringari, Sandro},
  journal = {Phys. Rev. A},
  volume = {87},
  issue = {3},
  pages = {033620},
  numpages = {3},
  year = {2013},
  month = {Mar},
  publisher = {American Physical Society},
  doi = {10.1103/PhysRevA.87.033620},
  url = {https://link.aps.org/doi/10.1103/PhysRevA.87.033620}
}

@misc{Fernandez2025,
      title={Angular momentum of rotating fermionic superfluids by {Sagnac} phonon interferometry}, 
      author={Marcia Frómeta Fernández and Diego Hernández Rajkov and Giulia Del Pace and Nicola Grani and Massimo Inguscio and Francesco Scazza and Sandro Stringari and Giacomo Roati},
      year={2025},
      eprint={2511.02664},
      archivePrefix={arXiv},
      url={https://arxiv.org/abs/2511.02664}, 
}

@article{Huang2025,
  title = {Emergence of Sound in a Tunable {Fermi} Fluid},
  author = {Huang, Songtao and Ji, Yunpeng and Repplinger, Thomas and Assump\c{c}\~ao, Gabriel G. T. and Chen, Jianyi and Schumacher, Grant L. and Vivanco, Franklin J. and Kurkjian, Hadrien and Navon, Nir},
  journal = {Phys. Rev. X},
  volume = {15},
  issue = {1},
  pages = {011074},
  numpages = {13},
  year = {2025},
  month = {Mar},
  publisher = {American Physical Society},
  doi = {10.1103/PhysRevX.15.011074},
  url = {https://link.aps.org/doi/10.1103/PhysRevX.15.011074}
}

@article{Hu2004,
  title = {Collective Modes and Ballistic Expansion of a {Fermi} Gas in the {BCS-BEC} Crossover},
  author = {Hu, Hui and Minguzzi, A. and Liu, Xia-Ji and Tosi, M. P.},
  journal = {Phys. Rev. Lett.},
  volume = {93},
  issue = {19},
  pages = {190403},
  numpages = {4},
  year = {2004},
  month = {Nov},
  publisher = {American Physical Society},
  doi = {10.1103/PhysRevLett.93.190403},
  url = {https://link.aps.org/doi/10.1103/PhysRevLett.93.190403}
}

@article{Schunck2008,
	Author = {Schunck, Christian H and Shin, Yong and Schirotzek, Andre and Ketterle, Wolfgang},
	Isbn = {0028-0836},
	Journal = {Nature},
	Number = {7205},
	Pages = {739--743},
	Publisher = {Macmillan Publishers Limited. All rights reserved},
	Title = {{Determination of the fermion pair size in a resonantly interacting superfluid}},
	Url = {http://dx.doi.org/10.1038/nature07176 http://www.nature.com/nature/journal/v454/n7205/suppinfo/nature07176{\_}S1.html},
	Volume = {454},
	Year = {2008}}

@misc{SI,
  note = {See Supplemental Material for details on experimental methods and Navier--Stokes simulations.}
}

@misc{Liao2026,
      title={Tunable viscosity across the {BCS-BEC} crossover}, 
      author={Yunxiang Liao and Andrey Grankin and Archisman Panigrahi and Victor Galitski and Leonid Levitov},
      year={2026},
      eprint={2604.10759},
      archivePrefix={arXiv},
      url={https://arxiv.org/abs/2604.10759}, 
}

@article{Krinner2017,
doi = {10.1088/1361-648X/aa74a1},
url = {https://doi.org/10.1088/1361-648X/aa74a1},
year = {2017},
month = {jul},
publisher = {IOP Publishing},
volume = {29},
number = {34},
pages = {343003},
author = {Krinner, Sebastian and Esslinger, Tilman and Brantut, Jean-Philippe},
title = {Two-terminal transport measurements with cold atoms},
journal = {Journal of Physics: Condensed Matter}
}

@article{Ketterle2008,
	Author = {Ketterle, W. and Zwierlein, M.W.},
	Issn = {0393697X},
	Journal = {Rivista del Nuovo Cimento},
	Number = {5-6},
	Title = {Making, probing and understanding ultracold {Fermi} gases},
	Volume = {31},
	Year = {2008},
}

@inproceedings{Zwierlein2016a,
	Author = {Zwierlein, M. W.},
	Booktitle = {Proceedings of the International School of Physics "Enrico Fermi"},
	Doi = {10.3254/978-1-61499-694-1-143},
	Isbn = {9781614996934},
	Issn = {18798195},
	Pages = {143--220},
	Publisher = {IOS Press},
	Title = {{Thermodynamics of strongly interacting Fermi gases}},
	Volume = {191},
	Year = {2016}}

@article{Schaefer2009,
	Author = {Thomas, Sch{\"{a}}fer and Derek, Teaney},
	Isbn = {0034-4885},
	Journal = {Reports on Progress in Physics},
	Number = {12},
	Pages = {126001},
	Title = {{Nearly perfect fluidity: from cold atomic gases to hot quark gluon plasmas}},
	Url = {http://stacks.iop.org/0034-4885/72/i=12/a=126001},
	Volume = {72},
	Year = {2009}}

@Inbook{Schaefer2012,
author="Sch{\"a}fer, T.
and Chafin, C.",
editor="Zwerger, Wilhelm",
title="Scaling Flows and Dissipation in the Dilute {Fermi} Gas at Unitarity",
bookTitle="The BCS-BEC Crossover and the Unitary Fermi Gas",
year="2012",
publisher="Springer Berlin Heidelberg",
address="Berlin, Heidelberg",
pages="375--406",
isbn="978-3-642-21978-8",
doi="10.1007/978-3-642-21978-8_10",
url="https://doi.org/10.1007/978-3-642-21978-8_10"
}

@article{Hou2021,
  title = {Dissipative superfluid hydrodynamics for the unitary {Fermi} gas},
  author = {Hou, Jiaxun and Sch\"afer, Thomas},
  journal = {Phys. Rev. A},
  volume = {104},
  issue = {2},
  pages = {023313},
  numpages = {12},
  year = {2021},
  month = {Aug},
  publisher = {American Physical Society},
  doi = {10.1103/PhysRevA.104.023313},
  url = {https://link.aps.org/doi/10.1103/PhysRevA.104.023313}
}

@article{Joseph2015,
  title = {Shear Viscosity of a Unitary {Fermi} Gas Near the Superfluid Phase Transition},
  author = {Joseph, J. A. and Elliott, E. and Thomas, J. E.},
  journal = {Phys. Rev. Lett.},
  volume = {115},
  issue = {2},
  pages = {020401},
  numpages = {5},
  year = {2015},
  month = {Jul},
  publisher = {American Physical Society},
  doi = {10.1103/PhysRevLett.115.020401},
  url = {https://link.aps.org/doi/10.1103/PhysRevLett.115.020401}
}

@article{Ku2012,
    author = {Mark J. H. Ku and Ariel T. Sommer and Lawrence W. Cheuk and Martin W. Zwierlein},
    title = {Revealing the Superfluid Lambda Transition in the Universal Thermodynamics of a Unitary {Fermi} Gas},
    journal = {Science},
    volume = {335},
    number = {6068},
    pages = {563-567},
    year = {2012},
    doi = {10.1126/science.1214987},
    url = {https://www.science.org/doi/abs/10.1126/science.1214987}
}

@article{Kuznetsov2020,
  title = {Expansion of the strongly interacting superfluid {Fermi} gas: Symmetries and self-similar regimes},
  author = {Kuznetsov, E. A. and Kagan, M. Yu. and Turlapov, A. V.},
  journal = {Phys. Rev. A},
  volume = {101},
  issue = {4},
  pages = {043612},
  numpages = {9},
  year = {2020},
  month = {Apr},
  publisher = {American Physical Society},
  doi = {10.1103/PhysRevA.101.043612},
  url = {https://link.aps.org/doi/10.1103/PhysRevA.101.043612}
}

@book{LandL,
    author = {L. D. Landau and E. M. Lifshitz}, 
    year = 1987,
    title = {Fluid Mechanics}, 
    edition = {Second},
    publisher = {Pergamon Press}
}

@article{Luciuk2017,
  title = {Observation of Quantum-Limited Spin Transport in Strongly Interacting Two-Dimensional {Fermi} Gases},
  author = {Luciuk, C. and Smale, S. and B\"ottcher, F. and Sharum, H. and Olsen, B. A. and Trotzky, S. and Enss, T. and Thywissen, J. H.},
  journal = {Phys. Rev. Lett.},
  volume = {118},
  issue = {13},
  pages = {130405},
  numpages = {6},
  year = {2017},
  month = {Mar},
  publisher = {American Physical Society},
  doi = {10.1103/PhysRevLett.118.130405},
  url = {https://link.aps.org/doi/10.1103/PhysRevLett.118.130405}
}

@article{Menotti2002,
  title = {Expansion of an Interacting {Fermi} Gas},
  author = {Menotti, C. and Pedri, P. and Stringari, S.},
  journal = {Phys. Rev. Lett.},
  volume = {89},
  issue = {25},
  pages = {250402},
  numpages = {4},
  year = {2002},
  month = {Dec},
  publisher = {American Physical Society},
  doi = {10.1103/PhysRevLett.89.250402},
  url = {https://link.aps.org/doi/10.1103/PhysRevLett.89.250402}
}

@book{Zwerger2012,
	editor = {Zwerger, Wilhelm},
	isbn = {3642219772},
	publisher = {Springer},
	title = {The BCS-BEC crossover and the unitary {Fermi} gas},
	volume = {836},
	year = {2012}}

@article{Li2024,
  title = {Universal density shift coefficients for the thermal conductivity and shear viscosity of a unitary {Fermi} gas},
  author = {Li, Xiang and Huang, J. and Thomas, J. E.},
  journal = {Phys. Rev. Res.},
  volume = {6},
  issue = {4},
  pages = {L042021},
  numpages = {5},
  year = {2024},
  month = {Oct},
  publisher = {American Physical Society},
  doi = {10.1103/PhysRevResearch.6.L042021},
  url = {https://link.aps.org/doi/10.1103/PhysRevResearch.6.L042021}
}

@article{Ketterle2008Varenna,
    author = {Wolfgang Ketterle and Martin W. Zwierlein}, 
    title = {Making, probing and understanding ultracold {Fermi} gases},
    address={IT},
   volume={31},
   ISSN={0393697X, 0393697X},
   url={https://doi.org/10.1393/ncr/i2008-10033-1},
   DOI={10.1393/ncr/i2008-10033-1},
   number={506},
   journal={La Rivista del Nuovo Cimento},
   publisher={SIF},
   year={2008},
   month=Jul, pages={247–422} }

@article{Giorgini2008,
	author = {Giorgini, Stefano and Pitaevskii, Lev P and Stringari, Sandro},
	journal = {Reviews of Modern Physics},
	number = {4},
	pages = {1215--1260},
	publisher = {APS},
	title = {{Theory of ultracold atomic Fermi gases}},
	url = {http://link.aps.org/abstract/RMP/v80/p1215},
	volume = {80},
	year = {2008}}

@phdthesis{PatelThesis,
  author  = {Patel, Parth},
  title   = "{Quantum transport in strongly interacting, ultracold {Fermi} gases in box potentials}",
  school  = {Massachusetts Institute of Technology},
  year    = "{2022}"
}

@book{Pitaevskii2016,
	address = {Oxford},
	author = {Pitaevskii, Lev P. and Stringari, Sandro},
	publisher = {Oxford University Press},
	title = {{Bose-Einstein Condensation and Superfluidity}},
	year = {2016}}

@article{Tey2013,
  title = {Collective Modes in a Unitary {Fermi} Gas across the Superfluid Phase Transition},
  author = {Tey, Meng Khoon and Sidorenkov, Leonid A. and Guajardo, Edmundo R. S\'anchez and Grimm, Rudolf and Ku, Mark J. H. and Zwierlein, Martin W. and Hou, Yan-Hua and Pitaevskii, Lev and Stringari, Sandro},
  journal = {Phys. Rev. Lett.},
  volume = {110},
  issue = {5},
  pages = {055303},
  numpages = {5},
  year = {2013},
  month = {Jan},
  publisher = {American Physical Society},
  doi = {10.1103/PhysRevLett.110.055303},
  url = {https://link.aps.org/doi/10.1103/PhysRevLett.110.055303}
}

@article{Joseph2011,
  title = {Observation of Shock Waves in a Strongly Interacting {Fermi} Gas},
  author = {Joseph, J. A. and Thomas, J. E. and Kulkarni, M. and Abanov, A. G.},
  journal = {Phys. Rev. Lett.},
  volume = {106},
  issue = {15},
  pages = {150401},
  numpages = {4},
  year = {2011},
  month = {Apr},
  publisher = {American Physical Society},
  doi = {10.1103/PhysRevLett.106.150401},
  url = {https://link.aps.org/doi/10.1103/PhysRevLett.106.150401}
}

@article{Xiang2024,
  title = {Universal density shift coefficients for the thermal conductivity and shear viscosity of a unitary {Fermi} gas},
  author = {Li, Xiang and Huang, J. and Thomas, J. E.},
  journal = {Phys. Rev. Res.},
  volume = {6},
  issue = {4},
  pages = {L042021},
  numpages = {5},
  year = {2024},
  month = {Oct},
  publisher = {American Physical Society},
  doi = {10.1103/PhysRevResearch.6.L042021},
  url = {https://link.aps.org/doi/10.1103/PhysRevResearch.6.L042021}
}

@article{
Xi2022,
author = {Xi Li  and Xiang Luo  and Shuai Wang  and Ke Xie  and Xiang-Pei Liu  and Hui Hu  and Yu-Ao Chen  and Xing-Can Yao  and Jian-Wei Pan },
title = {Second sound attenuation near quantum criticality},
journal = {Science},
volume = {375},
number = {6580},
pages = {528-533},
year = {2022},
doi = {10.1126/science.abi4480}}

@article{Shuryak2017,
  title = {Strongly coupled quark-gluon plasma in heavy ion collisions},
  author = {Shuryak, Edward},
  journal = {Rev. Mod. Phys.},
  volume = {89},
  issue = {3},
  pages = {035001},
  numpages = {61},
  year = {2017},
  month = {Jul},
  publisher = {American Physical Society},
  doi = {10.1103/RevModPhys.89.035001},
  url = {https://link.aps.org/doi/10.1103/RevModPhys.89.035001}
}

@Inbook{Schmitt2018,
author="Schmitt, Andreas
and Shternin, Peter",
editor="Rezzolla, Luciano
and Pizzochero, Pierre
and Jones, David Ian
and Rea, Nanda
and Vida{\~{n}}a, Isaac",
title="Reaction Rates and Transport in Neutron Stars",
bookTitle="The Physics and Astrophysics of Neutron Stars",
year="2018",
publisher="Springer International Publishing",
address="Cham",
pages="455--574",
isbn="978-3-319-97616-7",
doi="10.1007/978-3-319-97616-7_9",
url="https://doi.org/10.1007/978-3-319-97616-7_9"
}

@article{Keimer2015,
	author = {Keimer, B. and Kivelson, S. A. and Norman, M. R. and Uchida, S. and Zaanen, J.},
	da = {2015/02/01},
	doi = {10.1038/nature14165},
	id = {Keimer2015},
	isbn = {1476-4687},
	journal = {Nature},
	number = {7538},
	pages = {179--186},
	title = {From quantum matter to high-temperature superconductivity in copper oxides},
	ty = {JOUR},
	url = {https://doi.org/10.1038/nature14165},
	volume = {518},
	year = {2015}}

@article{Mukherjee2017,
  title = {Homogeneous Atomic {Fermi} Gases},
  author = {Mukherjee, Biswaroop and Yan, Zhenjie and Patel, Parth B. and Hadzibabic, Zoran and Yefsah, Tarik and Struck, Julian and Zwierlein, Martin W.},
  journal = {Phys. Rev. Lett.},
  volume = {118},
  issue = {12},
  pages = {123401},
  numpages = {5},
  year = {2017},
  month = {Mar},
  publisher = {American Physical Society},
  doi = {10.1103/PhysRevLett.118.123401},
  url = {https://link.aps.org/doi/10.1103/PhysRevLett.118.123401}
}

@article{Ohara2002,
    author = {K. M. O'Hara  and S. L. Hemmer  and M. E. Gehm  and S. R. Granade  and J. E. Thomas },
    title = {Observation of a Strongly Interacting Degenerate {Fermi} Gas of Atoms},
    journal = {Science},
    volume = {298},
    number = {5601},
    pages = {2179-2182},
    year = {2002},
    doi = {10.1126/science.1079107},
    url = {https://www.science.org/doi/abs/10.1126/science.1079107}
}

@article{Patel2020,
    author = {Parth B. Patel  and Zhenjie Yan  and Biswaroop Mukherjee  and Richard J. Fletcher  and Julian Struck  and Martin W. Zwierlein },
    title = {Universal sound diffusion in a strongly interacting {Fermi} gas},
    journal = {Science},
    volume = {370},
    number = {6521},
    pages = {1222-1226},
    year = {2020},
    doi = {10.1126/science.aaz5756},
    URL = {https://www.science.org/doi/abs/10.1126/science.aaz5756}
}

@article{Riemann1860, 
    author = {Bernhard Riemann},
    title = {{\"Uber} die {Fortpflanzung} ebener {Luftwellen} von endlicher {Schwingungsweite}},
    journal = {Abhandlungen der K\"oniglichen Gesellschaft der Wissenschaften zu G\"ottingen}, 
    volume = {8},
    year = {1860}
}

@article{Sommer2011,
  title={Universal spin transport in a strongly interacting {Fermi} gas},
  author={Sommer, Ariel and Ku, Mark and Roati, Giacomo and Zwierlein, Martin W},
  journal={Nature},
  volume={472},
  number={7342},
  pages={201--204},
  year={2011},
  publisher={Nature Publishing Group UK London}
}

@article{Wang2024,
  title = {Scale Invariance of a Spherical Unitary {Fermi} Gas},
  author = {Wang, Lu and Yan, Xiangchuan and Min, Jing and Sun, Dali and Xie, Xin and Peng, Shi-Guo and Zhan, Mingsheng and Jiang, Kaijun},
  journal = {Phys. Rev. Lett.},
  volume = {132},
  issue = {24},
  pages = {243403},
  numpages = {6},
  year = {2024},
  month = {Jun},
  publisher = {American Physical Society},
  doi = {10.1103/PhysRevLett.132.243403},
  url = {https://link.aps.org/doi/10.1103/PhysRevLett.132.243403}
}

@article{Yan2024,
author = {Zhenjie Yan  and Parth B. Patel  and Biswaroop Mukherjee  and Chris J. Vale  and Richard J. Fletcher  and Martin W. Zwierlein },
title = {Thermography of the superfluid transition in a strongly interacting {Fermi} gas},
journal = {Science},
volume = {383},
number = {6683},
pages = {629-633},
year = {2024},
doi = {10.1126/science.adg3430},
URL = {https://www.science.org/doi/abs/10.1126/science.adg3430}
}

@article{Zurn2013,
  title = {Precise Characterization of $^{6}\mathrm{Li}$ {Feshbach} Resonances Using Trap-Sideband-Resolved RF Spectroscopy of Weakly Bound Molecules},
  author = {Z\"urn, G. and Lompe, T. and Wenz, A. N. and Jochim, S. and Julienne, P. S. and Hutson, J. M.},
  journal = {Phys. Rev. Lett.},
  volume = {110},
  issue = {13},
  pages = {135301},
  numpages = {5},
  year = {2013},
  month = {Mar},
  publisher = {American Physical Society},
  doi = {10.1103/PhysRevLett.110.135301},
  url = {https://link.aps.org/doi/10.1103/PhysRevLett.110.135301}
}

\clearpage

%========================================
% Supplemental materials
%========================================
%TC:ignore
\setcounter{equation}{0}
\setcounter{figure}{0}
\setcounter{secnumdepth}{2}
\renewcommand{\theequation}{S\arabic{equation}}
\renewcommand{\thefigure}{S\arabic{figure}}
\renewcommand{\tocname}{Supplementary Materials}
\renewcommand{\appendixname}{Supplement}

% \tableofcontents
% \appendix

\section*{Supplementary Information}

\subsection{Polarization rotation imaging} 

Due to the high optical density of the atomic clouds (OD $\sim$ 10), we image the gas using a polarization rotation imaging technique described in detail in Ref.~\cite{PatelThesis}. Atom number calibration has been performed using both clock shifts in a weakly interacting $\ket{1}-\ket{3}$ mixture and measurements of the compressibility of a low temperature ideal Fermi gas, finding consistent results for the density to within $10\%$.

\subsection{Sound resonance thermometry}
For the unitary gas, the sound speed $c_0$ allows a determination of the per-particle energy via $\frac{E}{N} = \frac{9}{10} m c^2$~\cite{Patel2020}. From our measured atom number we obtain the Fermi energy $E_F$ and, using the equation of state~\cite{Ku2012}, we convert $\frac{E}{N E_F}$ to a temperature $\frac{T}{T_F}$, which we quote in Figure 3 of the main text.

\subsection{Deviation from self-similarity}
For the RMS deviation $\Delta \tilde{\rho}$ characterizing self-similarity in Fig.~4(c) of the main text, we combine data from all expansion times $t$ into the set $\{(\tilde{\xi}, \tilde{\rho})\}$. We bin the full range of $\tilde{\xi}$ values into $N = 30$ intervals, equally spaced in $\tilde{\xi}$. On each interval, we perform a local linear regression and estimate the variance of the points. Averaging the variances for each bin, weighted by the number of samples, we obtain an average variance and thus an average RMS deviation.

\subsection{Wall Sharpness}

The sharpness of the axially-confining endcap walls is inherently limited by the finite numerical aperture of the focusing lens.
This leads to a non-zero width of the transition region in which the atomic density within our box goes from $\rho_0$ to $0$. This finite width provides an initial length scale to the system which, for sufficiently small expansion time, limits self-similarity; for this reason we exclude profiles with $t < 1.0$ ms for the collapsed data.

\begin{figure}
    \includegraphics[width = 0.5\linewidth]{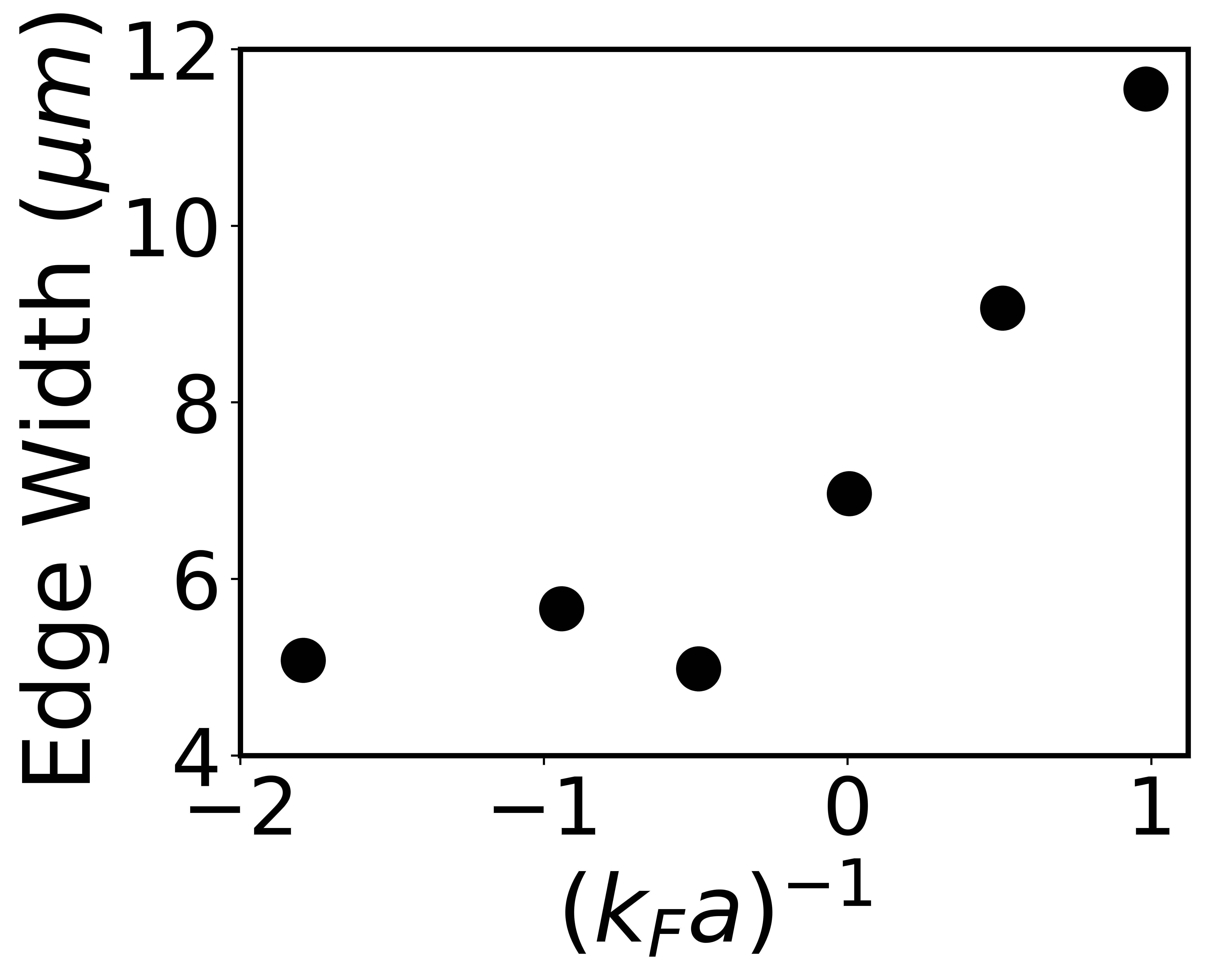}
    \caption{Box wall widths vs. $(k_F a)^{-1}$. The widths are from an error function fit ($\mathrm{erf}\left(\frac{x}{\sigma}\right)$) to the edges of the $t = 0$ ms profiles of the datasets shown in Fig. 4 of the main text.} 
    \label{Supp_Edge_Width}
\end{figure}

Fig.~\ref{Supp_Edge_Width} shows the edge width versus $\left(k_F a\right)^{-1}$ for the $t=0\,\rm ms$ data of Fig. 4 of the main text. In the BCS regime up to about unitarity, the width remains $\approx 5\,\rm \mu m$, but then grows to $\approx 10\,\rm \mu m$ in the BEC regime. The weakly interacting BEC has a smaller chemical potential than the Fermi gas at unitarity and the BCS regime~\cite{Ketterle2008Varenna} and thus probes the endcap wall at lower intensities, further in the wings of the laser beam.
The increased thickness of the density step may be contributing to the deviation from self-similarity for expansion on the BEC side (Fig. 4(c) of the main text), in addition to the effect of non-zero viscosity.

\subsection{Simulations} 

To investigate the role of viscosity on rarefaction waves and the influence of a finite wall width, we perform numerical simulations of the 1D expansion dynamics of a polytropic gas, using the full 1D Navier-Stokes equations: 

\begin{equation}
    \frac{\partial \rho}{\partial t} + v \frac{\partial \rho}{\partial x} + \rho \frac{\partial v}{\partial x} = 0
    \label{Supp_Continuity_Equation}
\end{equation}
\begin{equation}
    \frac{\partial v}{\partial t} + v \frac{\partial v}{\partial x} + \frac{1}{\rho} \frac{\partial}{\partial x} \left(P - \frac{4}{3} \eta \frac{\partial v}{\partial x}\right) = 0
    \label{Supp_Momentum_Equation}
\end{equation}
\begin{equation} 
\frac{\partial s}{\partial t} + v \frac{\partial s}{\partial x} - \frac{1}{\rho T} \left(\frac{4}{3} \eta \left(\frac{\partial v}{\partial x}\right)^2 + \frac{\partial}{\partial x} \left(\kappa \frac{\partial T}{\partial x}\right)\right) = 0
\label{Supp_Entropy_Equation}
\end{equation}

\begin{figure} 
    \includegraphics[width = \linewidth]{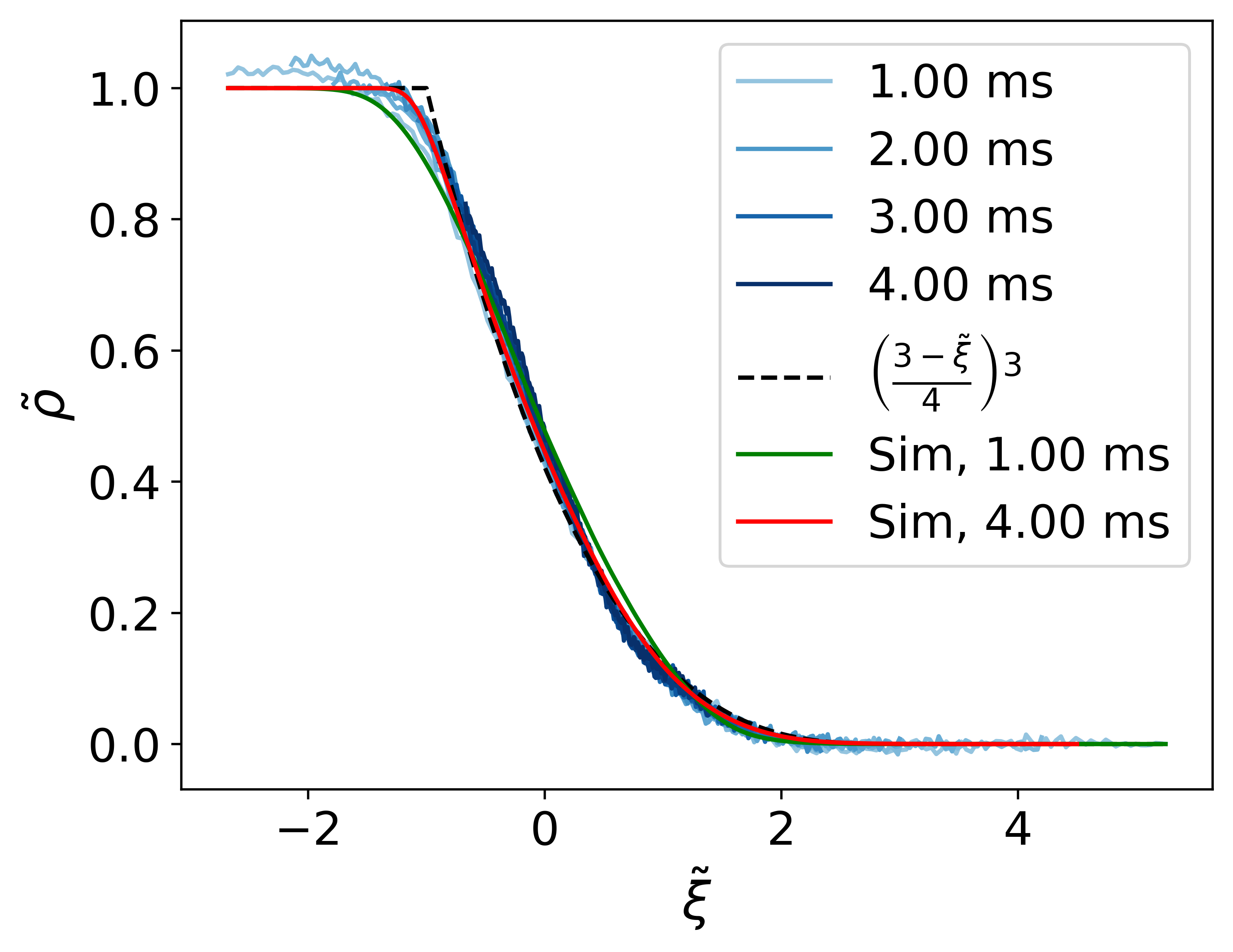}
    \caption{A comparison of the experimentally observed $\tilde{\rho}(\tilde{\xi})$ data with a 1D Navier stokes simulation. Blue curves (experimental data) and dashed black line (analytic prediction) are from Fig. 2 of the main text. Green and red curves are $\tilde{\rho}(\tilde{\xi})$ values obtained from a simulation whose parameters - initial wall width, speed of sound, sound diffusivity, etc. - were chosen to match the experimental data.}
    \label{Supp_Sim_Experiment_Comparison}
\end{figure}

Note that the bulk viscosity $\zeta$ is identically zero in the scale-invariant unitary gas, and is thus omitted. In simulations, the quantities $\rho, v, s$ are used as state variables; the quantities $P(\rho, s)$ and $T(\rho, s)$ are calculated using the equation of state given in Ref. \cite{Ku2012}. 

 As input to our simulations, we use an initial state where $v = 0$ everywhere, and where the spatial profile of $\rho$ is given by 

\begin{equation}
    \tilde{\rho}(x) = \frac{1 - \epsilon}{2} \left(1 - \mathrm{erf}\left(\frac{x}{\sigma}\right) \right) + \epsilon
    \label{Supp_Rho_Initial_Profile}
\end{equation}
Here $\epsilon$ is a small, unphysical density introduced to avoid division by zero in e.g. Eq. \ref{Supp_Momentum_Equation}; for presented simulations, $\epsilon = 10^{-6}$, which we have verified does not meaningfully affect our dynamics.

To initialize $s$, we postulate an initial reduced temperature $\theta_0 = \lim_{x \rightarrow -\infty} \frac{T}{T_F}$ at the peak density $\tilde{\rho} = 1$, then assume that the temperature $T$ is constant through the gas. The resulting temperature profile $\theta(x)$ is then converted to a an initial entropy profile $s(x)$ by using the equation of state from Ref. \cite{Ku2012}. For the presented simulations, $\frac{T}{T_F} = 0.15$, matching our lowest-temperature unitary data.

Other parameters were also chosen to match the experimental data. The simulated wall width $\sigma = 6.3\ \mu$m was chosen to match an average of the wall width for the $\ket{1}$ and $\ket{3}$ atoms  in the low-temperature data at unitarity. The background speed of sound $c_0 = 16.6\ \mu$m/ms is set to match Fig. 2 in the main text. The sound diffusivity measured for the unitarity data is insufficient to specify both the viscosity $\eta$ and the thermal conductivity $\kappa$; we initialize both by using the functional forms $\eta\left(\frac{T}{T_F}\right)$ and $\kappa\left(\frac{T}{T_F}\right)$ from Ref. ~\cite{Li2024}, at our initial temperature of $\frac{T}{T_F} = 0.15$; these chosen values imply a sound diffusivity which is within $10\%$ of the experimentally measured value. We make the simplifying assumption that $\nu$ and $\frac{\kappa}{\rho}$ remain constant throughout the gas, in order to avoid numerical instability due to a diverging viscosity in the low density tail.

To evolve the system, we convert Equations \ref{Supp_Continuity_Equation}, \ref{Supp_Momentum_Equation}, and \ref{Supp_Entropy_Equation} into a dimensionless form, with $c_0 = 1$ by construction, density normalized by $\rho_0$, and time units normalized by the arbitrary scale $t_0 = 1$ ms, then integrate the normalized equations numerically. We cross-check the numerical stability and error of our simulation by verifying that energy is conserved to at minimum the $10^{-4}$ level in all simulations presented. By comparing different step sizes, we verify that finite step size effects are negligible.

\begin{figure}[h]
    \includegraphics[width = \linewidth]{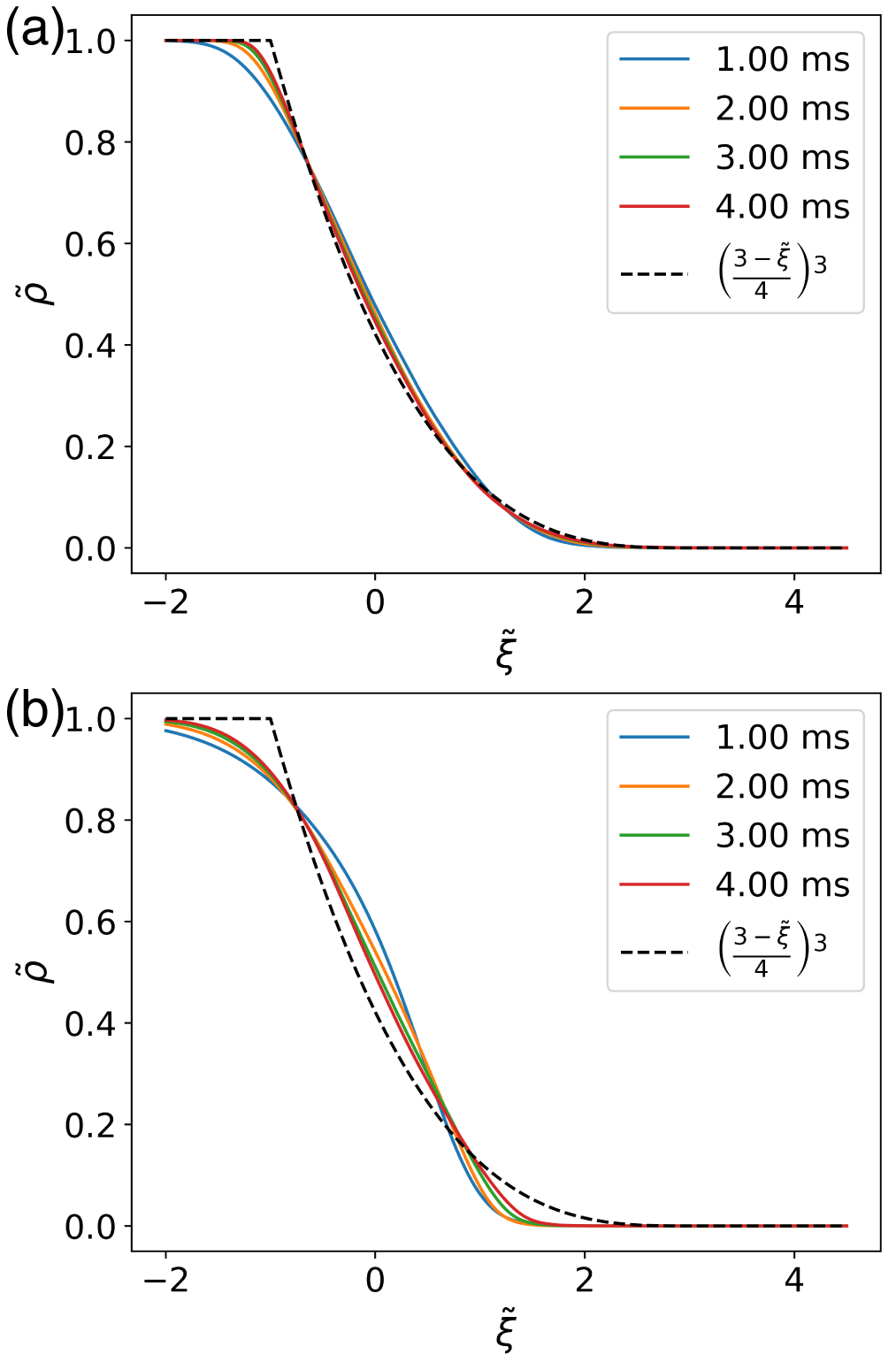}
    \caption{A comparison of simulations with different diffusivities. (a) A plot of the same simulated data as Fig. \ref{Supp_Sim_Experiment_Comparison}, with $\tilde{\rho}(\tilde{\xi})$ shown for different $t$. (b) Simulated data as (a), but with $\eta$ and $\kappa$ both multiplied by a factor of $20$.}
    \label{Supp_Sim_Viscosity_Comparison}
\end{figure}

In Figure \ref{Supp_Sim_Experiment_Comparison}, we present a comparison of the results of these simulations with experimental data, alongside the analytic prediction for the inviscid Euler dynamics.

From Fig. \ref{Supp_Sim_Experiment_Comparison}, we see that the simulations qualitatively agree with our experimental data - including the undershoot of the experimental data relative to the inviscid prediction near $\tilde{\xi} = -1$. This feature has contributions from both finite viscosity and the initially nonzero width of our simulated density profile - the qualitative agreement in simulation and experiment support that these known effects are sufficient to explain the deviations of our experimental data from the ideal profile.

Then, in Fig. \ref{Supp_Sim_Viscosity_Comparison}, we focus on the influence of nonzero viscosity. In Fig. \ref{Supp_Sim_Viscosity_Comparison} (a), we present a set of traces from the same simulations as Fig. \ref{Supp_Sim_Experiment_Comparison} at different times - as expected, we see that the simulated $\tilde{\rho}$ profiles converge to the analytic profile at late times. In Figure \ref{Supp_Sim_Viscosity_Comparison} (b), we present the results of simulations where $\eta$ and $\kappa$ have been multiplied by a factor of 20 compared to (a), but all other parameters are identical. We see that the most obvious effect of viscosity is to drive a deviation from the ideal profile as given by Eq. 9 of the main text. Also visible, but less dramatic, is a deviation from self-similarity, especially in the range $-0.5 \lesssim \tilde{\xi} \lesssim 0.5$. 

To approximately quantify these effects, we can apply the same metric as in Fig. 4(c) of the main text to probe the RMS deviation of the simulated datasets from an ideal self-similar curve. We find that the self-similar deviation is $4.6\times 10^{-3}$ for the low-diffusivity data and $9.4\times 10^{-3}$ for the high-diffusivity data. We then see that the increased $\eta$ and $\kappa$ \emph{do} decrease the self-similarity of the simulated data; however, the magnitude of this increase is smaller than the baseline deviation observed for the low-viscosity experimental data in Fig. 4(c) of the main text. Thus, the influence of viscosity on non-self-similarity in the deep BCS regime is likely being masked by experimental noise.

\end{document}